\documentclass{spie}
\usepackage{graphicx}
\def\etal{{\em et al.}}
\title{OPEN-LOOP WOOFER-TWEETER CONTROL ON THE LAO MULTI-CONJUGATE ADAPTIVE OPTICS TESTBED}

\author{EDWARD LAAG\supit{a}, DON GAVEL\supit{b} and MARK AMMONS\supit{b}
\skiplinehalf
\supit{a}Department of Earth Sciences, University of California Riverside, Riverside, CA 92521, USA; 
\supit{b}Laboratory for Adaptive Optics, 1156 High St., Santa Cruz, CA 95064, USA
}

\authorinfo{Further author information: (Send correspondence to E.A.L.)\\E-mail: edward.laag@ucr.edu}

\begin{document}

\maketitle

\begin{abstract}
Advances in micro deformable mirror (DM) technologies such as MEMs, have stimulated interest in the characteristics of systems that include a high stroke mirror in series with a high actuator count mirror.  This arrangement is referred to as a woofer-tweeter system.  In certain situations it may be desirable or necessary to operate the woofer DM in open-loop.  We present a simple method for controlling a woofer DM in open loop provided the device behaves in an approximately linear fashion.  We have tested a mirror that we believe meets our criterion, the ALPAO DM52 mirror.  Using our open-loop method we fit several test Kolmogorov wavefronts with the mirror and have achieved an accuracy of approximately 25 $nm$ $rms$ surface deviation over the whole clear aperture, and 20 $nm$ $rms$ over 90\% of the aperture.  We have also flattened the mirror in open loop to approximately 11 $nm$ $rms$ residual.
\end{abstract}

\section{Motivation}

\subsection{The MCAO Testbed}

The Lab for Adaptive Optics (LAO) currently has a testbed dedicated to the development of two key AO technologies for large telescopes (called multi-conjugate AO (MCAO)\cite{bec88} and multi-object AO (MOAO)).  Both of these technologies take advantage of tomographic reconstructions using multiple guidestars (a.k.a. reference sources)\cite{taf90}.  In particular, MCAO attempts to achieve a high strehl over a large field of view (FOV) by accounting for anisoplanatism, using multiple deformable mirrors at optical conjugates.  MOAO attempts to achieve very high strehls over small FOVs embedded in larger uncorrected fields.  First results from the testbed were shown in Ammons 2006\cite{amm06}.  Recent results have demonstrated the effectiveness of tomography at finding the layers ofturbulence, and high strehls have been achieved with both MCAO and MOAO.  

The testbed uses three optically addressed spatial light modulators (SLMs) from Hamamatsu Photonics.  The SLMs allow us to have nearly 600,000 control elements, far more than any current MEMs.  SLMs have been used with some success in the biological sciences (for example, a demonstration is described in Awwal 2003)\cite{aww03}.  Because their stroke is limited to approximately 1 wavelength deviation (about 650 $nm$ on the testbed) and we would like to avoid phase-wrapping, we are incorporating a high stroke mirror into the testbed both to eliminate the need for the SLMs to phase wrap, and to test possible AO configurations for future systems called ``woofer-tweeter'' setups.  

Taking an analogy from audio technology, the woofer-tweeter configuration in AO refers to the pairing of a higher resolution DM that has small stroke together with a high stroke (and consequently low resolution) DM called a woofer.  Though our testbed currently uses SLMs, woofer-tweeter combinations will also be useful for MEMs DMs. In addition to our lab, similar architectures are being studied at U. Victoria \cite{kes06} and at NUI Galway.

\begin{figure}
   \begin{center}
   \begin{tabular}{c}
   \includegraphics[height=9cm]{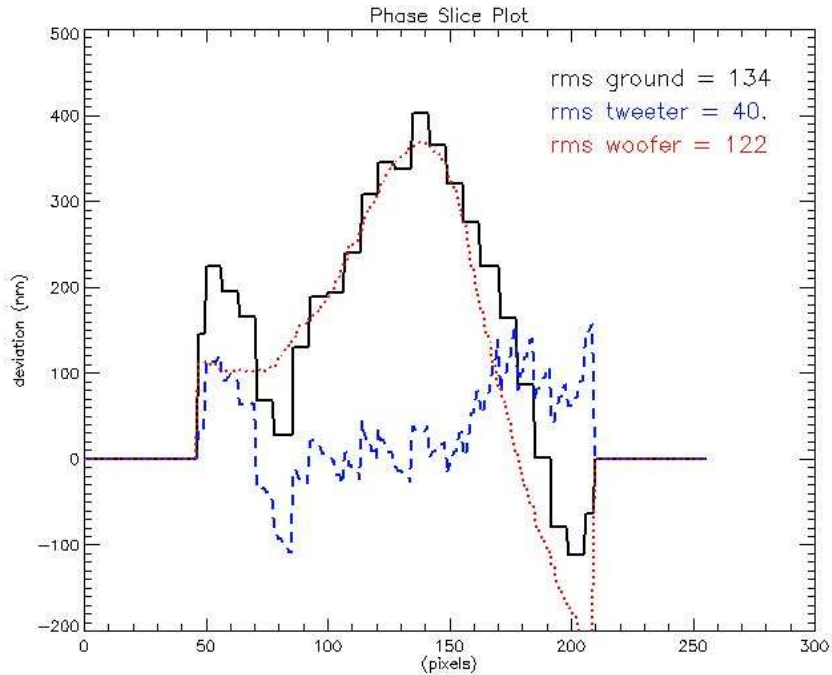}
   \end{tabular}
   \end{center}
   \caption{ \label{fig:fig1} 
A slice through the ground layer wavefront (shown as a solid line), with the woofer wavefront (dotted line) and the tweeter wavefront (dashed line) overlaid.}
   \end{figure}

\subsection{Design Considerations}

We created  a model of the ideal woofer DM by considering the influence functions to be an array of Gaussians that add linearly.  The ideal DM was configured with as few as 5 actuators across and as many as 9 across.  Using wavefronts measured on the testbed, we then tried to fit these shapes with our simulated DMs in software.  It was determined that a mirror of at least 6-7 actuators across was necessary to avoid phase wrapping which occurs any time peak-to-valley (P-V) aberrations exceed  650 $nm$ on our testbed.  Fig.~\ref{fig:fig1} shows an example of one of the simulations.  We looked into several mirror options on the market, starting with a small electrostatic device.  Due in part to the complex nature of the aberrations we are trying to correct, and the high predictability we need for open-loop performance, we eventually settled on an ALPAO DM52 mirror.  The importance of this high predictability (linearity) will be discussed below.

\begin{figure}
   \begin{center}
   \begin{tabular}{c}
   \includegraphics[height=9cm]{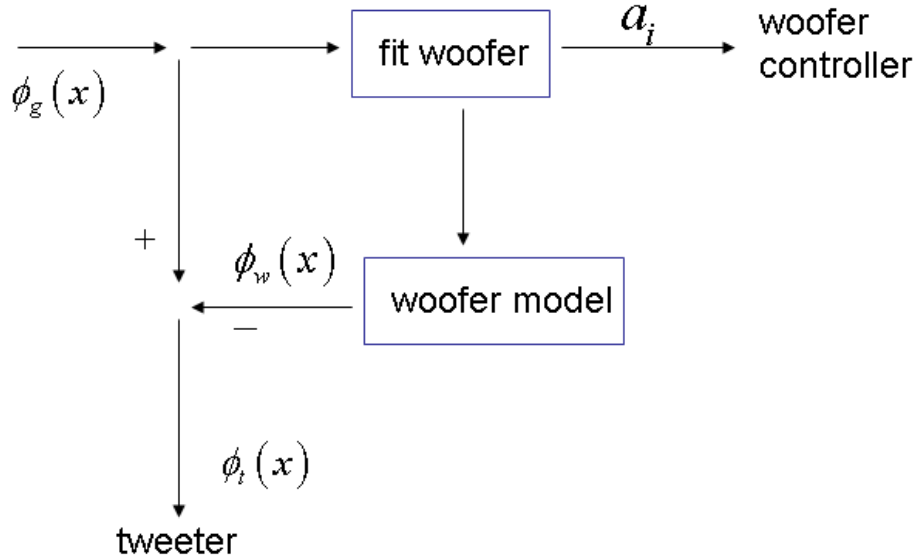}
   \end{tabular}
   \end{center}
   \caption{ \label{fig:fig2} 
Schematic diagram showing data flow for woofer-tweeter control.   The measured wavefront comes in at left.}
   \end{figure} 

\section{The ``Wooferfit'' Linear Super-Position Method}

We have put together some simple software for controlling a woofer-tweeter system in open loop (schematic shown in Fig.~\ref{fig:fig2}).  We call the routine that produces the open loop command signals for the mirror ``Wooferfit''.  Wooferfit runs after the tomography reconstructor has determined the turbulent layers in the volume. We put the low order components of the ground layer wavefront on the woofer.   

Wooferfit uses a simple linear super-position method to determine the commands which we will detail below.  This method makes the important assumption that the DM is approximately linear with input voltage commands and that the response functions superimpose linearly.   Obviously, due to hysteresis effects and force cross-coupling through the mirror face sheet, most DMs will not meet this requirement.  The ALPAO DM52 was designed with reduction of these effects in mind.  

The first step in our open-loop control process is to obtain good representations of the actuator influence functions using an interferometer.  Then, given a mirror with number of actuators $n$, the cross-talk matrix $R_{ij}$, is an $n$ by $n$ sized array generated from: 

\begin{equation}
\label{eq:eq1}
R_{ij} = \int r_{i}(x)\cdot r_{j}(x)dx
\end{equation}

where $r_{i}(x)$ are the infuence functions over the mirror surface $x$, previously measured when a unit of voltage is applied to actuator $i$.  The voltage commands $a_{i}$ to the DM controller device are simply:

\begin{equation}
\label{eq:eq2}
a_{i} = R_{inv}\cdot \int r_{i}(x) \cdot \phi_{g}(x)dx
\end{equation}

and the wavefront will be given by:

\begin{equation}
\label{eq:eq3}
\phi_{w}(x) = \sum_{i} a_{i}\cdot r_{i}(x)
\end{equation}

\section{Results}

\begin{figure}
   \begin{center}
   \begin{tabular}{c}
   \includegraphics[height=11cm]{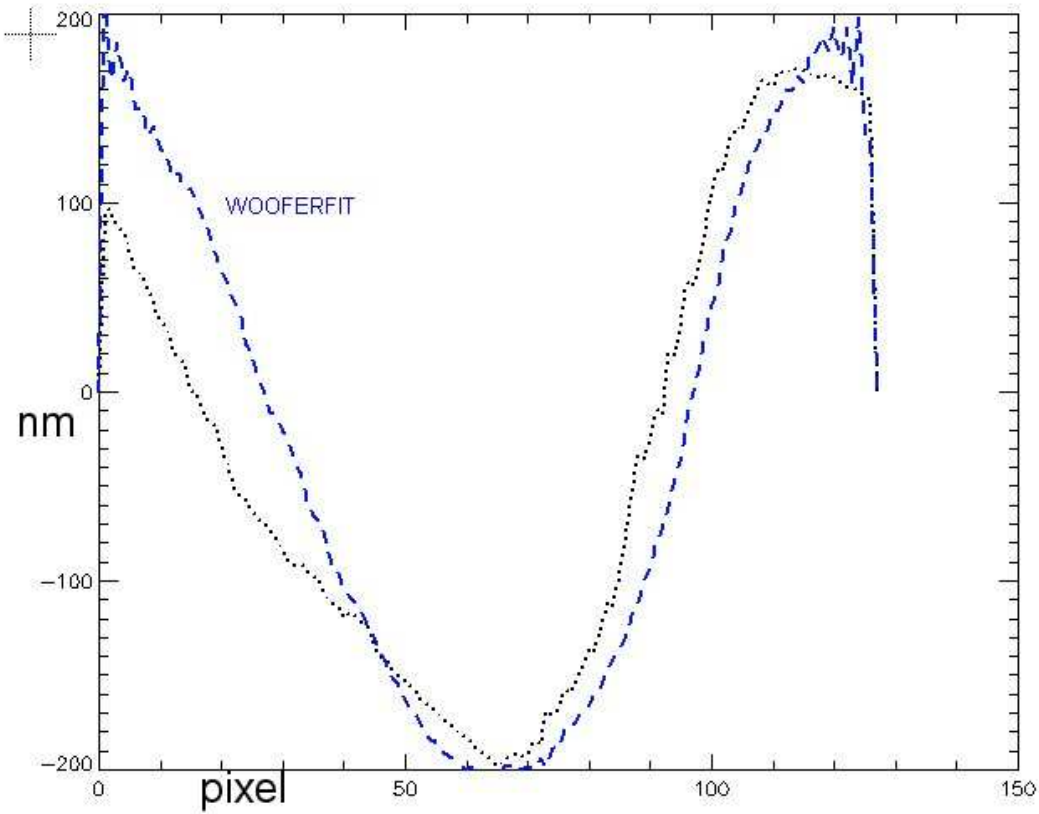}
   \end{tabular}
   \end{center}
   \caption{ \label{fig:fig3} 
A lineout of a random row from the comparison of the predicted wooferfit WF (dashed line) and the measured Zygo wavefront (dotted line).}
   \end{figure}

The particular ALPAO DM52 mirror we have in the lab has about a 133 $nm$ $rms$ focus shape when initially powered on but with no commands sent.  In order to generate a flat shape we measured and inverted this wavefront and ran it through Wooferfit to generate flattening commands.  Our first attempt at flattening the DM52 in open-loop resulted in a residual of approximately 11 $nm$ $rms$ of surface flatness deviation over the full clear aperture of the mirror. 

We then tried to fit a typical Kolmogorov wavefront.  The testbed uses etched glass Kolmogorov phase plates as turbulence generators.  The phase plates are meant to simulate a normal atmosphere's worth of wavefront aberration.  We measured the wavefront using a set of Shack-Hartmann wavefront sensors.  After doing a tomographic reconstruction of the estimated volume, there is a residual on the ground layer with approximately 250 $nm$ $rms$ tip/tilt removed wavefront error.  We then fit this Kolmogorov wavefront with the ALPAO DM52 using Wooferfit. We compared the surface of the mirror as measured by a Zygo interferometer to the wavefront generated by the tomography software.  Our comparison shows a 25 $nm$ $rms$ disagreement between the ALPAO DM52 and the Wooferfit predicted shape over the clear aperture (see Fig.~\ref{fig:fig3} above).  The fit was noticeably better within the central portion of the mirror and when apertured down to 90\% of the clear aperture the agreement was roughly 20 $nm$ $rms$.

It is important to note that these results are significant because they represent open-loop “go-to” control of the surface without the benefit of feedback from residual wavefront measurements.  Hence these results are applicable to systems which need to run open-loop like MOAO configurations mentioned earlier.

\section{Conclusion}

We have tested the suitability of the ALPAO DM52 as a woofer DM and have shown it has promising open-loop characteristics.  Initial results look good for woofer-tweeter implementation in our MCAO testbed.

\acknowledgements

The authors are grateful for funding from the Gordon \& Betty Moore foundation, whose generous gift enabled the construction of the Laboratory for Adaptive Optics at the University of CA, Santa Cruz.  The authors also acknowledge the generous support of the Bachmann family.  This work has been supported by the National Science Foundation Science and Technology Center for Adaptive Optics, managed by the University of California at Santa Cruz under cooperative agreement No. AST - 9876783.  

Special thanks to David Anderson from the Herzberg Institute of Astrophysics who graciously supplied me with influence functions for the ALPAO mirror.

\end{document}